\documentclass[12pt,epsf]{article}
\usepackage{verbatim}
\usepackage{anyfontsize}
\usepackage{dsfont}
\usepackage{tikz}
\usepackage{sidecap}
\sidecaptionvpos{figure}{t}
\usepackage{subfigure}

\usepackage[english]{babel}
\usepackage{enumitem}  

\usepackage{amssymb,amsmath}
\usepackage{graphicx, xcolor, varwidth}
\usepackage{setspace}
\usepackage[permil]{overpic}
\usepackage{cite}
\usepackage[square, comma, sort&compress,numbers]{natbib}

\DeclareSymbolFont{matha}{OML}{txmi}{m}{it}
\DeclareMathSymbol{v}{\mathord}{matha}{118}

\colorlet{darkblue}{blue!70!black}
\colorlet{darkgreen}{green!70!black}

\usepackage[colorlinks=true,urlcolor=darkblue,linktocpage=true,linkcolor=darkblue,citecolor=darkblue]{hyperref}

\numberwithin{equation}{section}
\DeclareMathSymbol{v}{\mathord}{matha}{118}
\newcommand{\be}{\begin{equation}}
\newcommand{\ee}{\end{equation}}
\newcommand{\bea}{\begin{eqnarray}}
\newcommand{\eea}{\end{eqnarray}}
\newcommand{\bear}{\begin{eqnarray}}
\newcommand{\eear}{\end{eqnarray}}
\newcommand{\beas}{\begin{eqnarray*}}

\newcommand{\eeas}{\end{eqnarray*}}
\newcommand{\ba}{\begin{array}}
\newcommand{\ea}{\end{array}}

\newcommand{\tr}{\operatorname{tr}}
\newcommand{\pd}[2][1]{\ifnum#1=1 \frac{\partial}{\partial {#2}} \else
  \frac{\partial^#1}{\partial {#2}^{#1}}\fi}
\newcommand{\dpd}[2][1]{\ifnum#1=1 \dfrac{\partial}{\partial {#2}} \else
  \frac{\partial^#1}{\partial {#2}^{#1}}\fi}
\newcommand{\td}[2][1]{\ifnum#1=1 \frac{d}{d{#2}} \else
  \frac{d^#1}{d{#2}^{#1}}\fi}

\def\H{{\cal H} }

\renewcommand{\(}{\left(}
\renewcommand{\)}{\right)}

\newcommand{\nbox}{{\,\lower0.9pt\vbox{\hrule \hbox{\vrule height 0.2 cm \hskip 0.19 cm \vrule height 0.2 cm}\hrule}\,}}
\newcommand{\Tr}{\ {\rm Tr}\ }

\def\O{{\cal O}}



\begin{document}
\begin{spacing}{1.3}
\begin{titlepage}

\begin{center}
{\Large \bf 

Bulk Matter and the Boundary Quantum Null Energy Condition

}

\vspace*{6mm}

Zuhair U. Khandker$^a$, Sandipan Kundu$^b$, Daliang Li$^b$

\vspace*{6mm}

\textit{$^{a}$Department of Physics, University of Illinois, 1110 W. Green St., Urbana IL 61801-3080, USA\\}

\vspace{3mm}

\textit{$^{b}$Department of Physics and Astronomy,  Johns Hopkins University, Charles Street, Baltimore, MD 21218, USA\\}

\vspace{6mm}

{\tt \small zuhair@illinois.edu, kundu@jhu.edu, daliang.li@jhu.edu}

\vspace*{6mm}
\end{center}

\begin{abstract}
We investigate the quantum null energy condition (QNEC) in holographic CFTs, focusing on half-spaces and particular classes of states. We present direct, and in certain cases nonperturbative, calculations for both the diagonal and off-diagonal variational derivatives of entanglement entropy. In $d\geq3$, we find that the QNEC is saturated. We compute relations between the off-diagonal variation of entanglement, boundary relative entropy, and the bulk stress tensor. Strong subadditivity then leads to energy conditions in the bulk. In $d=2$, we find that the QNEC is in general not saturated when the Ryu-Takayanagi surface intersects bulk matter. Moreover, when bulk matter is present the QNEC can imply new bulk energy conditions. For a simple class of states, we derive an example that is stronger than the bulk averaged null energy condition and reduces to it in certain limits.
\end{abstract}

\end{titlepage}
\end{spacing}

\vskip 1cm
\setcounter{tocdepth}{2}  
\tableofcontents

\begin{spacing}{1.3}

\section{Introduction and summary}

In relativistic QFTs, the quantum null energy condition (QNEC) provides a striking connection between energy and entanglement. First proposed in~\cite{Bousso:2015mna}, the QNEC states that at any point $p$, the expectation value of the null-null component of the stress tensor, in any state, is bounded from below by a certain derivative  (defined below) of entanglement entropy:
\be
\langle T_{--}(p) \rangle \geq \frac{1}{2\pi} S^{\prime\prime}.
\label{QNEC1}
\ee
Proofs of this statement currently exist for free and superrenormalizable theories~\cite{Bousso:2015wca}, holographic theories~\cite{Koeller:2015qmn}, and completely general interacting theories in $d\geq 3$ spacetime dimensions~\cite{Balakrishnan:2017bjg}, all of which have established the QNEC as a fascinating new energy condition in QFT.\footnote{In this paper, we only consider QFTs in Minkowski space. For work on the QNEC in curved space, see for example~\cite{Fu:2017evt,Akers:2017ttv,Fu:2017lps,Fu:2017ifb}.}  In this note, we study the QNEC for some simple holographic states, focusing on relationships between bulk matter and boundary entanglement.  

The right-hand side of (\ref{QNEC1}) is defined as follows. With the point $p$ lying on some Cauchy surface, consider an entangling cut that runs through $p$ and divides space into two regions. Then $S$ is the von Neumann entropy associated with this division. In particular, describing the entangling cut using embedding functions $x^\mu(y)$, for some internal coordinate $y$, the entropy is a functional $S=S[x^\mu(y)]$. The second variation of $S$ in the null direction will have both a diagonal and off-diagonal piece,
\be
\frac{\delta^2 S}{\delta x^-(y)\delta x^-(y^\prime)} = S^{\prime\prime}(y)\delta(y-y^\prime) + S_{\mathrm{od}}(y,y^\prime),
\label{SVar}
\ee
where ``od" stands for ``off-diagonal". What appears in (\ref{QNEC1}) is the diagonal piece $S^{\prime\prime}$.  

One goal of this work is to study $S^{\prime\prime}$ and $S_{\mathrm{od}}$ in simple holographic settings, where these quantities can be computed explicitly (to leading order in $N$) using the Ryu-Takayanagi (RT) prescription, and to draw lessons from the concrete calculations. In $d\geq 3$ spacetime dimensions, we focus on two classes of CFT states: (i) \emph{nonperturbative} states described by the restricted metric
\be\label{nonpert}
ds^2= 
\frac{1}{z^2}(-dx^+ dx^- + d \vec{y}^2 +  dz^2) + h_{--}\left( x^-,\vec{y},z \right)  (dx^-)^2\ ,
\ee
where $(x^\pm,\vec{y})$ are boundary coordinates, $z$ is the bulk coordinate, and $h_{--}$ is arbitrary (this includes the important example of shockwaves); and (ii) \emph{perturbative} states described by the general metric
\be\label{pert}
ds^2= 
\frac{1}{z^2}(-dx^+ dx^- + d \vec{y}^2 +  dz^2) + h_{\mu \nu}\left(x^+, x^-,\vec{y},z \right) dx^\mu dx^\nu\ ,
\ee
with $h_{\mu\nu}$ small. For \emph{both} of these classes of states, with a half-space entangling cut, we obtain
\begin{eqnarray}
S^{\prime\prime} &=& 2\pi \langle T_{--} \rangle \label{S''} \\[5pt]
S_{\mathrm{od}}(\vec{y}_1,\vec{y}_2) &=&-2\pi\int \frac{dz d\vec{y}}{z^{d-1}} \, g(z,\vec{y};\vec{y}_1)g(z,\vec{y};\vec{y}_2) T_{--}^{bulk}, \label{Sod}
\end{eqnarray}
where $g(z,\vec{y};\vec{y}^\prime)$ is the Green's function defined in (\ref{g}). 

These results allow us to draw the following conclusions in $d\geq 3$ for the nonperturbative $h_{--}$ states in (\ref{nonpert}) and the perturbative $h_{\mu\nu}$ states in (\ref{pert}):
\begin{itemize}[leftmargin=5mm,topsep=5pt]
\itemsep0em
\item[1.] The QNEC is saturated (eq. \ref{S''}).\footnote{In \cite{Ecker:2017jdw}, the authors used numerical methods to probe a weak version of the QNEC (which includes contributions from $S_{\mathrm{od}}$) for colliding shockwave states. They obtained evidence that the weak QNEC can be saturated in complicated excited states in holographic theories.}
\item[2.] Strong subadditivity (SSA) implies energy conditions in the bulk. In particular it is known that SSA implies $S_{\mathrm{od}}\leq 0$~\cite{Bousso:2015mna}. This immediately puts a sign constraint on the integral of the bulk stress tensor appearing on the right-hand side of (\ref{Sod}). Moreover, for the states (\ref{nonpert}), we will show that the full implication is actually the boundary null energy condition. That is, the states (\ref{nonpert}) satisfy $\langle T_{--} \rangle \geq 0$ on the boundary.
\item[3.] Boundary relative entropy is related to $S_{\mathrm{od}}$ and can also be written as an integral of $T_{--}^{bulk}$ (eq. \ref{SrelTbulk}). We note that the first law of entanglement~\cite{Blanco:2013joa}, $\delta S = \delta \langle K \rangle$ where $K$ is the modular Hamiltonian, holds for linear perturbations of the density matrix. At this order, $S_{\mathrm{od}}$ and relative entropy vanish. From (\ref{Sod}), we see explicitly that metric perturbations induced by bulk matter generally correspond to second-order density-matrix perturbations on the boundary.
\end{itemize}

We also consider $d=2$ spacetime dimensions. In 2d, there is no notion of $S_{\mathrm{od}}$, since there is no $\vec{y}$-direction. It has been proven within holography that the 2d QNEC takes a stronger form~\cite{Koeller:2015qmn},
\be\label{qnec2}
 \langle T_{--} \rangle \ge \frac{\hbar}{2\pi}\left(S'' + \frac{6}{c}S'^2\right)\ ,
\ee 
where $c$ is the central charge. We show that: 
\begin{itemize}[leftmargin=5mm,topsep=5pt]
\itemsep0em
\item[4.] The QNEC is saturated in holographic CFT$_2$ for any boundary interval in any state with a geometric dual if the corresponding RT surface does not intersect bulk matter. 
\item[5.] The 2d QNEC is generally \emph{unsaturated} when the RT surface passes through bulk matter. This is a stark difference from the higher dimensional case detailed above, where saturation occurs even in presence of bulk matter. In particular, for perturbative states of the form (\ref{nonpert}) we find
\be\label{2dQNECNon}\nonumber
\left(S^{\prime\prime}+\frac{6}{c} S^{\prime2}\right)(x^{-}_2)=2\pi T_{--}(x^{-}_2)-\pi\int_{x^-_1}^{x^-_2} dx^- \frac{\left(x^--x^-_1\right)^3 }{\left(x^-_2-x^-\right) \left(x^-_2-x^-_1\right)^3 }T^{bulk}_{--}(z(x^-),x^-),
\ee
where the integral is along a bulk geodesic.  For these states the 2d QNEC implies the energy condition (\ref{anec2}) on the bulk stress tensor, which is weaker than the null energy condition but stronger than the averaged null energy condition. This energy condition is different from ones derived using strong subadditivity, \emph{e.g.}, ~\cite{Lashkari:2014kda,Bhattacharya:2014vja}.  
\end{itemize} 

\textit{Note: While this manuscript was in preparation, \cite{Leichenauer:2018obf} has appeared on arXiv with a remarkable result: every state in a holographic CFT in $d\ge3$ saturates the QNEC. The authors start from the JLMS result~\cite{Jafferis:2015del} that boundary relative entropy equals bulk relative entropy and then use general arguments to show that (\ref{S''}) should hold universally in holography. By comparison, we have not assumed JLMS in our calculations, and so our derivation of (\ref{S''}), especially for the nonperturbative $h_{--}$ states, should be seen as an explicit and independent check of the more general result of~\cite{Leichenauer:2018obf}, and also of JLMS. Our results in $CFT_2$ are new.}

\section{Preliminaries}
In quantum field theory, entanglement entropy is the von Neumann entropy associated with a region. Suppose we have a quantum system in state $\rho$ and we divide the total system into two subsystems $A$ and $\bar{A}$; the total Hilbert space is a direct product of two spaces $\H_{tot}=\H_A \otimes\H_{\bar{A}}$. The reduced density matrix for the region $A$ is obtained by tracing out degrees of freedom inside the subsystem $\bar{A}$: $\rho_A=\tr_{\bar{A}} \rho$. Given a reduced density matrix $\rho_A$, we can define a global operator:
\be
K_A(\rho)= -{\mathbf 1}_{\bar{A}}\otimes \ln \rho_A
\ee
which is known as the modular Hamiltonian. In general, the modular Hamiltonian is a complicated state dependent non-local operator which contains all the information about the region $A$. The entanglement entropy which is defined as
\be
S=\tr\left(\rho_A K_A(\rho) \right)
\ee
measures the amount of information loss due to tracing out the region $\bar{A}$. Thus, the entanglement entropy depends both on the quantum state of the system and the entangling surface. 

For CFTs that are dual to Einstein gravity, a precise prescription for computing entanglement entropy was proposed in \cite{Ryu:2006bv} and later generalized in \cite{Hubeny:2007xt}. According to the proposal, the entanglement entropy for a region $A$ is given by\footnote{In this work we only consider coherent states in the bulk, such that $S_{bulk}=0$ (see~\cite{Faulkner:2013ana}).}
\begin{equation}\label{ee}
S=\frac{\text{Area}\left(\gamma_A\right)}{4 G_N}, 
\end{equation}
where $G_N$ is the $(d+1)$-dimensional Newton's constant. $\gamma_A$ is the $(d-1)$-dimensional minimal area surface in the bulk whose boundary is given by the boundary of the region $A$: $\partial \gamma_A=\partial A$. The area of the surface $\gamma_A$ is denoted by $\text{Area}\left(\gamma_A\right)$.

In this note, we focus on the case where the region $A$ is a half-space. We denote coordinates on the $d$-dimensional boundary CFT as $(x^+,x^-,\vec{y})$,
where $x^\pm=x^0 \pm x^1$. The half-space is
\be
A = \left\{x^0=0, x^1>0 \right\}\ .
\ee

Now we briefly review the concept of relative entropy. Letting $\rho_0$ denote the vacuum state of a system and $\rho$ some other generic state, the relative entropy between $\rho$ and $\rho_0$ is defined to be
\be \label{Srel}
S(\rho||\rho_0) \equiv  \Delta \langle K_0  \rangle - \Delta S\ .
\ee
Here $K_0 = K_A (\rho_0)$ is the vacuum modular Hamiltonian, $\Delta \langle K_0  \rangle = \Tr (\Delta \rho K_0)$, and $\Delta S = S - S_0$. Relative entropy is a measure of the distinguishability of $\rho$ and $\rho_0$.

\section{QNEC in $d\ge 3$}

In this section, we derive (\ref{S''})-(\ref{Sod}) for both the nonperturbative $h_{--}$ states described by the metric (\ref{nonpert}) (sec \ref{subsec:nonpert}) and the perturbative $h_{\mu\nu}$ states described by the metric (\ref{pert}) (sec \ref{subsec:pert}), and then discuss the implications of these results (sec \ref{subsec:discuss}). 

\subsection{Nonperturbative $h_{--}$ states}
\label{subsec:nonpert}
In this subsection we analyze the states of finite bulk gravitational shockwaves. In particular, we are interested in states with the following metric:
\be\label{metric}
ds^2= 
\frac{1}{z^2}(-dx^+ dx^- + d \vec{y}^2 +  dz^2) + h_{--}\left( x^-,\vec{y},z \right)  (dx^-)^2\ ,
\ee
where $h_{--}$ is not necessarily small. It satisfies the Einstein's equation:
\be \label{Einstein--}
-\frac{1}{2}  \left(z^{d-1} \partial_z\frac{1}{z^{(d-1)}} \partial_z +\partial_{\vec{y}}^2 \right)z^2 h_{--}=8\pi G_N T_{--}^{bulk}\ ,
\ee
and has a Fefferman-Graham expansion near the boundary of AdS,
\be\label{stress}
h_{--}= \frac{16\pi G_N z^{d-2}}{d}\langle T_{--}\rangle+\cdots \ .
\ee
Here $\langle T_{\mu \nu}\rangle$ is the expectation value of the CFT stress tensor in the state dual to the geometry (\ref{metric}). An example of this kind of state is the planar delta function shockwave states studied in \cite{Afkhami-Jeddi:2017rmx} which can be created by inserting heavy operators on the boundary.

We introduce coordinates $X^\pm(z,\vec{y})$ to parametrize the RT surface. In pure AdS and for a half-space entangling cut, the RT surface has a simple profile: ${X^\pm}(z,\vec{y})=0$. Moreover, metric perturbations in the form (\ref{metric}) leaves this extremal surface invariant. One can check  this by deriving equations of motion from the RT action and then plugging in ${X^\pm}(z,\vec{y})=0$.

The variation of $S$ in (\ref{SVar}) is a measure of how entanglement entropy changes as a result of null deformations to the entangling surface. We choose a region $A$ to be a deformed half-space,
\be\label{half}
A=\left\{x^0=\frac{1}{2}\delta x^-(\vec{y}), \,\,\, x^1>-\frac{1}{2}\delta x^-(\vec{y}) \right\}\ ,
\ee
where $\delta x^-(\vec{y})$ is a positive function. For infinitesimal $\delta x^-(\vec{y})$, the corresponding deformation of the RT surface is also infinitesimal, and it is sufficient for us to expand the RT action perturbatively in $\delta X^{\pm}(z,\vec{y})$. The linear order RT action vanishes, indicating that ${X^{\pm}}(z,\vec{y})=0$ is indeed extremal in (\ref{metric}). The non-trivial contribution to the RT action appears at second order
\begin{align}\label{defRT2}
S_{RT}=\frac{1}{8G_N} &\int \frac{dz d^{d-2}\vec{y}}{z^{d-1}}\delta^{ab}\left[\partial_a (\delta X^-) \partial_b (\delta X^+)-z^2 h_{--}\partial_{a} (\delta X^-)\partial_{b} (\delta X^-)\right]\ ,
\end{align}
where the indices $a,b$ run over $\{z, \vec{y}\}$. Note that we have not assumed $h_{--}$ is small. 

The equations of motion following from (\ref{defRT2}) are
\begin{align}\label{eom}
&\delta^{ab} \partial_a \left(\frac{1}{z^{d-1}} \partial_b \right)\delta X^+-2\delta^{ab} \partial_a \left(\frac{h_{--}}{z^{d-3}} \partial_b \right)\delta X^-=0\ ,\nonumber\\
&\delta^{ab} \partial_a \left(\frac{1}{z^{d-1}} \partial_b \right)\delta X^-=0\ ,
\end{align}
with the boundary conditions $\delta X^+(z=0,\vec{y})=0$ and $\delta X^-(z=0,\vec{y})=\delta x^-(\vec{y})$.

The solution for $\delta X^-$ is
\be \label{Xm}
\delta X^-(z,\vec{y})= \int d^{d-2} \vec{y}' \delta x^-(\vec{y}') g(z,\vec{y};\vec{y}')  \ ,
\ee
where
\be \label{g}
g(z,\vec{y},\vec{y}') \equiv \frac{2^{d-2}\Gamma \left(\frac{d-1}{2}\right)}{ \pi ^{\frac{d-1}{2}}}  \frac{z^d}{\left( z^2+(\vec{y}-\vec{y}')^2 \right)^{d-1}} 
\ee
coincides with the bulk to boundary propagator.

For $\delta X^+$, it is convenient to write
\be \label{Xp}
\delta X^+(z,\vec{y})=z^2 h_{--}(z,\vec{y}) \delta X^-(z,\vec{y})+H(z,\vec{y})\ ,
\ee
such that $H(z,\vec{y})$ satisfies the differential equation
\begin{eqnarray}
\label{eqnH}
\delta^{ab} \partial_a \left(\frac{1}{z^{d-1}} \partial_b \right)H = -\delta X^-\delta^{ab} \partial_a \left(\frac{1}{z^{d-1}} \partial_b \right)z^2 h_{--} 
= 16\pi G_N \frac{1}{z^{d-1}}  \delta X^- T_{--}^{bulk}\ ,
\end{eqnarray}
with boundary condition $H(0,\vec{y})=0$, and we used Einstein's equation for the second equality. The solution is
\be\label{H}
H(z,\vec{y})= 16\pi G_N \int \frac{dz' d^{d-2}\vec{y}'}{z^{\prime d-1} } G(z,\vec{y};z',\vec{y}') \delta X^-(z^\prime, \vec{y}^\prime) T_{--}^{bulk}(z^\prime,\vec{y}^\prime)   \ ,
\ee
where $G(z,\vec{y};z',\vec{y}')$ is given by
\begin{align}
G(z,\vec{y};z',\vec{y}')&=-\frac{\Gamma\left[\frac{d+1}{2} \right]z z' (4\pi)^{\frac{1-d}{2}}}{d(d-1)}\left(\frac{\rho^2}{1-\rho^2} \right)^{1-d}\nonumber\\
&\hspace{5mm} \times {}_2F_1 \left(d-1,\frac{d+1}{2}, d+1,1-\frac{1}{\rho^2}\right)
\end{align}
with
\be
\rho=\sqrt{\frac{(z-z')^2+(\vec{y}-\vec{y}')^2}{(z+z')^2+(\vec{y}-\vec{y}')^2}}\ .
\ee
Note that
\be
- \frac{1}{z^{d-1}} \partial_z G(z,\vec{y};z',\vec{y}')\Big{|}_{z=\epsilon}=g(z',\vec{y}';\vec{y})\ .
\ee 

Substituting these solutions into the RT action and integrating by parts, we can write down the entanglement entropy as an integral on the boundary $z=\epsilon$:
\begin{align}
S =\frac{1}{16G_N}\int_{z=\epsilon}\frac{d^{d-2}\vec{y}}{z^{d-1}}&\left[\left(\delta X^-\right)^2\partial_z (z^2 h_{--})+\partial_z \left(\delta X^- H\right) \right]\ , \label{EE2}
\end{align}
Recall that $\delta X^-(z=0,\vec{y}) = \delta x^-(\vec{y})$. Hence from (\ref{EE2}) we can read off the second variation of $S$ defined in (\ref{SVar}) and determine $S^{\prime\prime}$ and $S_{\mathrm{od}}$. The first term in (\ref{EE2}) contributes to $S^{\prime\prime}$ while the second term contributes to $S_{\mathrm{od}}$. Using the Fefferman-Graham expansion (\ref{stress}) and the asymptotic form of $H$ in (\ref{EE2}) we obtain
\begin{eqnarray} 
S'' &=& \frac{1}{8G_N}\left[\frac{\partial_z (z^2 h_{--})}{z^{d-1}}\right]_{z=\epsilon}=2\pi \langle T_{--}\rangle \label{S''2} \ , \\
S_{\mathrm{od}}(\vec{y}_1,\vec{y}_2) &=& -2\pi\int \frac{dz d^{d-2}\vec{y}}{z^{d-1}} g(z,\vec{y};\vec{y}_1)g(z,\vec{y};\vec{y}_2)  T_{--}^{bulk} \label{Sod2} \ .
\end{eqnarray}
This completes our derivation for the $h_{--}$ states. We see explicitly that the QNEC is saturated. We discuss this result further in sec (\ref{subsec:discuss}).

\subsection{Perturbative $h_{\mu\nu}$ states}
\label{subsec:pert}
Now, we outline the analogue calculation for the perturbative $h_{\mu\nu}\ll1$ states,
\be\label{metrich}
ds^2= 
\frac{1}{z^2}(-dx^+ dx^- +  dz^2+d \vec{y}^2) + h_{\mu \nu}dx^\mu dx^\nu\ ,
\ee
where $h_{\mu \nu}$ satisfies the linearized Einstein's equation in the bulk with a Fefferman-Graham expansion near the boundary of AdS
\be\label{stressh}
h_{\mu \nu}= \frac{16\pi G_N z^{d-2}}{d}\langle T_{\mu \nu}\rangle+\cdots \ .
\ee
The geometry (\ref{metrich}) is dual to CFT states that are perturbatively close to the vacuum with stress tensor $\langle T_{\mu \nu}\rangle$. Again we pick the region $A$ to be a deformed half-space (\ref{half}). 

Now the second order RT functional is given by
\begin{align}\label{defRT3}
S_{RT} =\frac{1}{8G_N} &\int_{X^\pm=0} \frac{dz d^2\vec{y}}{z^{d-1}}\delta^{ab}\Big[-\partial_a (\delta X^-) \partial_b (\delta X^+)+z^2 h_{--}\partial_{a} (\delta X^-)\partial_{b} (\delta X^-)\nonumber\\
& -z^{d-1}(\delta X^-)^2\partial_a \left(\frac{1}{z^{d-1}} \partial_- \right)z^2 h_{-b}+\frac{z^2}{2}(\delta X^-)^2 \partial_-^2 h_{ab}\Big]+\O(h^2)\ ,
\end{align}
where the indices $a,b$ run over $\{z, \vec{y}\}$.\footnote{We have been using the Fefferman-Graham gauge: $h_{\mu z}=0$.} The equations of motion are somewhat more involved now. However, it turns out that to linear order in $h_{\mu\nu}$, and utilizing Einstein's equations, the solutions for $\delta X^\pm(z,\vec{y})$ look exactly the same as the $h_{--}$ case considered above. In particular, the $\delta X^\pm(z,\vec{y})$ are again given by (\ref{Xm}), (\ref{Xp}), and (\ref{H}). Moreover, plugging these solutions into the action (\ref{defRT3}), it is straightforward to check that (\ref{S''2})-(\ref{Sod2}) again hold.

\subsection{Discussion}
\label{subsec:discuss}

Thus far, we have derived (\ref{S''})-(\ref{Sod}) for both the nonperturbative $h_{--}$ states (\ref{nonpert}) and the perturbative $h_{\mu\nu}$ states (\ref{pert}). We now discuss various lessons that can be drawn from these results. \\

\noindent \textbf{QNEC saturation}

Equation (\ref{S''2}) demonstrates that the QNEC is \emph{saturated}. This is striking because the null energy, which depend only on the degrees of freedom at a particular location in spacetime, is directly determined by the non-local entanglement entropy. For holographic CFTs with geometric bulk dual, integrating (\ref{S''2}) and (\ref{Sod2}) may provide a shortcut for computing the entanglement entropy from some simpler initial state.  

Note that (\ref{S''2}) holds regardless of whether or not bulk matter is present. This will no longer be true for holographic CFTs in $d=2$ spacetime dimensions. We will see in the next section that $T^{bulk}_{--}$ can spoil the QNEC saturation in $d=2$. \\

\noindent \textbf{Bulk energy condition}

It is known that strong subadditivity implies $S_{\mathrm{od}}(\vec{y}_1,\vec{y}_2) \leq 0$~\cite{Bousso:2015mna}. Thus, (\ref{Sod2}) immediately places an sign constraint on the integrated bulk stress tensor,
\be
\int \frac{dz d\vec{y}}{z^{d-1}} \, g(z,\vec{y};\vec{y}_1)g(z,\vec{y};\vec{y}_2) T_{--}^{bulk}(z,\vec{y}) \geq 0 \ . 
\ee
This is a new type of energy condition in the bulk where the null energy $T^{bulk}_{--}$ is averaged over the RT surface, which in this case is a $d-1$ dimensional spacelike plane. \\

\noindent \textbf{Null energy condition}

Interestingly, for the $h_{--}$ states, $S_{\mathrm{od}}(\vec{y}_1,\vec{y}_2) \leq 0$ actually implies the null energy condition (NEC) on the boundary. To see this, we first rewrite (\ref{Sod2}) using Einstein's equation (\ref{Einstein--}),
\be
S_{\mathrm{od}}(\vec{y}_1,\vec{y}_2)= \frac{1}{8G_N}\int dz d^{d-2}\vec{y} \, g(z,\vec{y};\vec{y}_1)g(z,\vec{y};\vec{y}_2) \delta^{ab} \partial_a \left(\frac{1}{z^{(d-1)}} \partial_b \right)z^2 h_{--}\ .
\ee
Note that unlike the diagonal piece $S''$, $S_{\mathrm{od}}$ depends on $h_{--}$ on the entire extremal surface.  However, the integral of $S_{\mathrm{od}}$ is again determined solely by the boundary behavior of the metric. More precisely, we integrate the equation above with respect to $\vec{y}_2$ and integrate by parts. This yields a boundary term at $z=\epsilon$ that can be evaluated using the Fefferman-Graham expansion (\ref{stress}) for $h_{--}$. The result is
\begin{eqnarray}
\int d^{d-2}\vec{y}_2 S_{\mathrm{od}}(\vec{y}_1,\vec{y}_2) &=& -\frac{1}{8G_{N}}\int_{z=\epsilon} d^{d-2}\vec{y} \, g\left(z,\vec{y},\vec{y}_{1}\right)\left(\frac{1}{z^{d-1}}\partial_{z}\right)z^{2}h_{--} \nonumber \\[10pt]
 &=& -2\pi \langle T_{--}\left(\vec{y}_{1}\right) \rangle \ . \label{intSod}
\end{eqnarray}
This result, along with the non-positivity of $S_{\mathrm{od}}$, implies the boundary NEC for the $h_{--}$ states,
\be
\langle T_{--}\rangle \ge 0\ .
\ee
Moreover, since we also have QNEC saturation $S^{\prime\prime} = 2\pi \langle T_{--}\rangle$, it follows that 
\be \label{Stot0}
\int d^{d-2}\vec{y}_2 \frac{\delta^2 S}{\delta x^-(\vec{y}_1)\delta x^-(\vec{y}_2)} = S^{\prime\prime}(\vec{y}_1) + \int d^{d-2}\vec{y}_2 S_{\mathrm{od}}(\vec{y}_1,\vec{y}_2)   = 0\ . 
\ee 
\vspace{2pt}

\noindent \textbf{Relative entropy and the first law}

Our results demonstrate the interesting relation between bulk matter and boundary relative entropy~\cite{Lin:2014hva,Jafferis:2015del}.\footnote{See also~\cite{Lashkari:2016idm,Neuenfeld:2018dim}.} Recall that the definition (\ref{Srel}) of relative entropy is $S(\rho||\rho_0) = \Delta \langle K_0  \rangle - \Delta S$. A key fact is that for deformed half-spaces, there exists an explicit expression for the vacuum subtracted modular energy $\Delta \langle K_0 \rangle$~\cite{Casini:2017roe} (see also~\cite{Faulkner:2015csl,Faulkner:2016mzt,Koeller:2017njr}), given by
\be \label{Casini}
\Delta \langle K_0 \rangle = 2\pi \int d^{d-2} \vec{y} \int_{\delta x^-(\vec{y})}^{\infty}dx^- \left(x^- - \delta x^-(\vec{y}) \right)\langle T_{--}(x^-,\vec{y}) \rangle \ .
\ee
Expanding this expression in $\delta x^-(\vec{y})$, it follows that 
\be
\frac{\delta^2 \langle K_0 \rangle}{\delta x^-(\vec{y}_1) \delta x^-(\vec{y}_2)} = 2\pi \langle T_{--}(\vec{y_1}) \rangle \delta(\vec{y}_1-\vec{y}_2) \ .
\ee
Then combining this with our results (\ref{S''})-(\ref{Sod}) for $S$, we immediately get that
\be\label{sholo}
\frac{\delta^2 S(\rho||\rho_0)}{\delta x^-(\vec{y}_1) \delta x^-(\vec{y}_2)}=-S_{\mathrm{od}}(\vec{y}_1,\vec{y}_2)=2\pi\int \frac{dz d^{d-2}\vec{y}}{z^{d-1}}g(z,\vec{y};\vec{y}_1)g(z,\vec{y};\vec{y}_2) T_{--}^{bulk}(z,\vec{y}) 
\ee
so the part of relative entropy quadratic in $\delta x^-$ is 
\begin{eqnarray}
S^{(2)}(\rho||\rho_0) 
&=& \pi\int \frac{dz d^{d-2}\vec{y} d^{d-2}\vec{y}_1 d^{d-2}\vec{y}_2}{z^{d-1}}g(z,\vec{y};\vec{y}_1)g(z,\vec{y};\vec{y}_2) T_{--}^{bulk}(z,\vec{y}) \delta x^-(\vec{y}_1) \delta x^-(\vec{y}_2) \nonumber \\
&=& \pi \int \frac{dz d^{d-2}\vec{y}}{z^{d-1}} \delta X^-(z,\vec{y})\delta X^-(z,\vec{y})T_{--}^{bulk}(z,\vec{y}) \ , \label{SrelTbulk}
\end{eqnarray}
for both the $h_{--}$ and perturbative $h_{\mu\nu}$ states we have considered. 

Generically $S(\rho||\rho_0) \neq 0$, even for the perturbative $h_{\mu\nu}$ states. Because the first law~\cite{Blanco:2013joa} $\delta S = \delta \langle K_0 \rangle$ must hold at first order in perturbations of the density matrix, this non-trivial relative entropy should appear at higher orders.\footnote{In~\cite{Casini:2017roe}, the author derive $\Delta K$ using the first law and $\Delta S$. However they noted that after taking a limit corresponding to spacelike entangling regions, 
$\Delta S \neq \Delta \langle K_0 \rangle$.} For the holographic states we are considering, we see from (\ref{sholo}) that the relative entropy is intimately related to the bulk stress tensor. 
Therefore metric perturbations induced by the bulk stress tensor generally correspond to second order perturbations of the density matrix. This is also consistent with the fact that the first law implies linearized Einstein's equation without source \cite{Faulkner:2013ica}, while the bulk source only appears after the second order effects are taken into account~\cite{Faulkner:2017tkh}.

\section{QNEC in $d=2$}
In 2d, the QNEC takes a somewhat different form~\cite{Koeller:2015qmn}:
\be
T_{--}\ge\frac{1}{2\pi}\left(S^{\prime\prime}+\frac{6}{c}S^{\prime2}\right)\ .
\label{eq:2dQNEC}
\ee
In this section, we investigate saturation of the QNEC in 2d holographic CFTs. Given a state in a holographic CFT, we show that the QNEC is saturated for any boundary interval whose corresponding RT surface goes across vacuum in the bulk. Then we show that if the RT surface passes through bulk matter, the QNEC is no longer saturated. For states perturbatively close to empty AdS, the deviation is proportional to the bulk stress tensor. The QNEC then implies an energy condition for matter in the bulk. 

\subsection{Saturation in ``vacuum-like" states}

We provide a general argument for saturation of the QNEC in the absence of bulk matter. The result applies to a wide family of bulk states beyond the empty AdS. In particular, we only require that the bulk stress tensor $T_{bulk}=0$ in a finite region containing the RT surface, which is a geodesic in $AdS_3$. We also present two explicit examples, a microstate of the BTZ blackhole and a shockwave with source deep in the bulk.

This general argument is based on three simple facts. The first is that for a simply connected bulk region that does not contain matter, there exist an uniformizing transformation that trivializes the metric in this region. 
The second is that the quantity $(S^{\prime\prime}+\frac{6}{c}S^{\prime2})$ transforms exactly like the boundary holomorphic stress tensor in 2d~\cite{Wall:2011kb}. The third is that the QNEC is saturated in the vacuum for any interval. All these elements were understood in various contexts in the literature. We will present a unified treatment in this section.

Our primary tool is the following relation between the renormalized entanglement entropy and a 2-pt function of primary operators:\footnote{The renormalization procedure corresponds to regularizing the geodesic length by inserting a UV brane $\epsilon$-away from the boundary and then subtracting divergent terms. This procedure does not affect $S^{\prime}$.}
\be
S=\frac{c}{6}\ell=-\frac{c}{12}\frac{1}{h}\log\langle O(x^{-}_{f},x^{+}_{f})O(x^{-}_{i},x^{+}_{i})\rangle_{s},
\label{eq:HoloS}
\ee
where $\ell=\frac{L}{L_{AdS}}$ is the renormalized geodesic length in units of the AdS scale. $O$ is any scalar Virasoro primary with weight $(h,h)$ satisfying $c\gg h\gg1$. We use the subscript $s$ to denotes the state. To derive this relation, we used the 2d version of the Ryu-Takayanagi formula \cite{Ryu:2006bv} which relates $S$ to the geodesic length $S=\frac{L}{4G}$, the relation between the UV renormalized geodesic length and a 2-pf of a heavy scalar operator, $\langle OO\rangle=e^{-2h L/L_{AdS}}$, as well as the Brown-Henneaux relation $c=\frac{3L_{AdS}}{2G}$ \cite{Brown:1986nw}.

We first test that the QNEC is saturated in the vacuum, where $T_{--}=0$. For any interval in the vacuum, we use (\ref{eq:HoloS}) to find $S=\frac{c}{6}\log x^{-}_{fi}$ where $x^{-}_{fi}\equiv x^{-}_{f}-x^{-}_{i}$. It is straight-forward to show that the RHS of (\ref{eq:2dQNEC}) is zero as well. 

Next, we show that the LHS and the RHS of (\ref{eq:2dQNEC}) transforms in the same way. Under a Virasoro transformation $x^{-}\rightarrow w^{-}(x^{-})$, $T_{--}$ transforms as: 
\be
\langle T_{--}(x^{-})\rangle_{s}=\left(\frac{dw^{-}}{dx^{-}}\right)^{2}\langle T_{--}(w^-)\rangle_{s^{\prime}}-\frac{1}{2\pi}\frac{c}{12}\{w^{-}(x^{-}),x^-\}
\ee
where $\{w^-,x^-\}$ is the Schwarzian derivative defined as:
\be
\{w^-(x^-),x^-\}\equiv\frac{w^{-\prime\prime\prime}w^{-\prime}-\frac{3}{2}w^{-\prime\prime2}}{w^{-\prime2}}
\ee
Using the transformation of a primary operator $O(x^-)= ({w^{-}}')^{h} O(w^-)$ in (\ref{eq:HoloS}), we find that the RHS of (\ref{eq:2dQNEC}) transforms in the exact same way:
\be
\frac{d^{2}S_{s}}{dx^{-2}_{f}}+\frac{6}{c}\left(\frac{dS_{s}}{dx^{-}_{f}}\right)^{2}=\left(\frac{dw^{-}_{f}}{dx^{-}_{f}}\right)^{2}\left(\frac{d^{2}S_{s^{\prime}}}{dw_{f}^{-2}}+\frac{6}{c}\left(\frac{dS_{s^{\prime}}}{dw^{-}_{f}}\right)^{2}\right)-\frac{c}{12}\{w^{-}_{f},x^{-}_{f}\}
\ee
Thus QNEC is saturated in any state that can be uniformized by a Virasoro transformation. 

We will provide two explicit examples demonstrating this general principle: a shockwave with finite energy and a state generated by a heavy operator insertion, which may be dual to a deficit angle geometry or a micro-state of a BTZ blackhole in the bulk. \\

\noindent \textbf{Shockwave with no bulk matter}

We first consider a shockwave geometry with the bulk metric given by: 
\be
ds^2=\frac{1}{z^2}(dx^{-} dx^{+}+dz^2)+A\delta(x^-)dx^{-2}
\ee
Note that there is no bulk source at finite $z$. This geometry can be understood as a limiting case of \cite{Afkhami-Jeddi:2017rmx} where the bulk source generating the shockwave is sent to $z\rightarrow\infty$. The boundary stress tensor is:
\be
T_{--}=\frac{c}{12\pi}A\delta(x^-).
\ee
To test whether the QNEC is saturated, we observe after the coordinate transformation
\be
w^-(x^-)=x^{-}-\Theta(x^-)\frac{x^{-2}A}{1+x^-A},\hspace{1cm}w^+=x^+,\hspace{1cm}u=z\sqrt{w^{-\prime}(x^-)},
\ee 
that the metric takes the form of the empty AdS metric in $(w^-,w^+,u)$ coordinates. We can compute the CFT 2-point function in the $(x^-,x^+)$ coordinate:
\be
\langle O(x^{-}_f,x^{+}_f) O(x^{-}_i,x^{+}_i) \rangle_{s}=\(\frac{dw^{-}_f}{dx^{-}_f}\)^{h}\(\frac{dw^{-}_i}{dx^{-}_i}\)^{h} \(\frac{dw^{+}_f}{dx^{+}_f}\)^{\bar{h}}\(\frac{dw^{+}_i}{dx^{+}_i}\)^{\bar{h}}\frac{1}{w^{-2h}_{fi}w_{fi}^{+2\bar{h}}}.
\ee
Plugging this into (\ref{eq:HoloS}), we get precisely: 
\be
\frac{1}{2\pi}\left(S^{\prime\prime}+\frac{6}{c}S^{\prime2}\right)=\frac{c}{12\pi}A\delta(x^-).
\ee
Thus the QNEC is saturated for a finite energy shockwave when the RT surface does not intersect any bulk source. This fact extends to finite-width shockwaves as well. \\

\noindent \textbf{Blackhole microstates and deficit angles}

We can also study the QNEC in states generated by a local operator insertions. We are particularly interested in a microstate of the BTZ blackhole generated by acting a heavy operators $O_H$ with $h_H \sim c$ on the vacuum. To compute the expectation value in this state, we insert one $O_H$ at the origin and another at $\infty$. In this state, we have:
\be
T_{--}=\frac{-1}{2\pi}\frac{h_{H}}{x^{-2}}.
\ee
The Virasoro transformation that normalizes this state is
\be
w^-\left(x^-\right)=x^{-\alpha},\hspace{1cm}\alpha=\sqrt{1-\frac{24h_{H}}{c}}
\ee
Note the transition at $h_H=\frac{c}{24}$. When $c\rightarrow\infty$, $h_H>\frac{c}{24}$ corresponds to a microstate of a BTZ blackhole, while $0<h_H<\frac{c}{24}$ corresponds to a heavy massive point particle generating an angle defect. 
In the large $c$ limit, we have \cite{Fitzpatrick:2014vua}:
\be
\langle O\left(x^{-}_{f},x^{+}_{f}\right)O\left(1,1\right)\rangle_{H}=\left(\frac{\alpha (x_{f}^{-})^{\frac{\alpha-1}{2}}}{1-(x_{f}^{-})^{\alpha}}\right)^{2h}\left(\frac{\alpha (x_{f}^{+})^{\frac{\alpha-1}{2}}}{1-(x_{f}^{+})^{ \alpha}}\right)^{2h}
\ee
Plugging into (\ref{eq:HoloS}), we get precisely: 
\be
\frac{1}{2\pi}\left(S^{\prime\prime}+\frac{6}{c}S^{\prime2}\right)=\frac{-1}{2\pi}\frac{h_{H}}{x^{-2}}.
\ee
Thus the QNEC is saturated. Note the RT surface does not intersect the singularity.

\subsection{Bulk matter and non-saturation}
In this section, we demonstrate that in 2d CFT, the QNEC is not saturated if the RT geodesic passes through bulk matter. We show this by analyzing a smeared shockwave with small energy on an empty $AdS_3$ background. The corresponding metric is:
\be\label{metric3}
ds^2= 
\frac{1}{z^2}(-dx^+ dx^- +  dz^2) + h_{--}(x^-,z)dx^- dx^-\ ,
\ee
where $h_{--}$ satisfies Einstein's equation in the bulk:
\be
-\frac{z^2}{2}\left( \partial_z^2 h_{--} +\frac{3}{z}\partial_z h_{--}\right)=8\pi G_N T_{--}^{bulk}(x^-,z) \ ,
\ee
where $T_{--}^{bulk}(x^-,z)$ is the bulk matter source. The expectation value of the CFT stress tensor in the state dual to the geometry (\ref{metric3}) is given by the asymptotic behavior of $h_{--}$:
\be\label{stress3}
\langle T_{--}(x^-)\rangle= \frac{h_{--}(x^-,z=0)}{ 8\pi G_N} \ .
\ee
Now, consider a spacelike interval on the boundary with end points: $(x^+_1,x^-_1)$ and $(x^+_2, x^-_2)$, where $x^-_2>x^-_1$. At the linear order in small $h_{--}$, the entanglement entropy is given by:
\be\label{length}
S =S^0+\frac{1}{4G_N}\int_{0}^1 d\lambda\ \lambda(1-\lambda)(x^-_2-x^-_1)^2  h_{--}(x^-(\lambda),z(\lambda))+ \O(h^2)\ ,
\ee
where, $S^0$ is the entanglement entropy in empty AdS and the geodesic is parametrized by:
\be
x^-(\lambda)=x^-_1+(x^-_2-x^-_1)\lambda\ , \qquad z(\lambda)=\sqrt{\lambda(1-\lambda)(x^-_2-x^-_1)(x^+_1-x^+_2) }\ .
\ee
This allows us to evaluate the right hand side of (\ref{eq:2dQNEC}) in the linear order in $h_{--}$:
\be 
\frac{d^2 S}{d {x^-_2}^2}+\frac{6}{c} \left(\frac{d S}{d {x^-_2}}\right)^2=\frac{d^2 S^1}{d {x^-_2}^2}+\frac{12}{c} \left(\frac{d S^0}{d {x^-_2}}\right)\left(\frac{d S^1}{d {x^-_2}}\right)+ \O(h^2)\ ,
\ee
where, $S^1$ is the $\O(h)$ term of (\ref{length}). In the last equation, the zeroth order terms drop out because as discussed in the last section, in empty AdS, entanglement entropy saturates the QNEC. We now take the derivates of equation (\ref{length}) and after some manipulation we obtain
\begin{align}\label{eqn12}
\frac{d^2 S}{d {x^-_2}^2}+\frac{6}{c} \left(\frac{d S}{d {x^-_2}}\right)^2=\int_{x^-_1}^{x^-_2} dx^- \frac{\left(x^--x^-_1\right) \left(x^--x^-_2\right)}{16 G_N \left(x^-_1-x^-_2\right)^3 } \Big[ 24 +24 \left(x^--x^-_1\right) \partial_-  \nonumber\\
\left. +11 z  \partial_z +z^2  \partial_z^2 +4 z \left(x^--x^-_1\right) \partial_z \partial_- + 4\left(x^--x^-_1\right)^2 \partial_-^2 \right]h_{--}(z,x^-)\ ,
\end{align}
where, now 
\be
z(x^-)=\sqrt{\frac{(x^+_1-x^+_2)\left(x^--x^-_1\right) \left(x^--x^-_2\right)}{(x^-_2-x^-_1)} }\ .
\ee
In the equation (\ref{eqn12}), we have set $L_{AdS}=1$ and used the fact that $c=\frac{3}{2G_N}$. After performing several integrations by parts in (\ref{eqn12}), we obtain
\begin{align}\label{eqn13}
\frac{d^2 S}{d {x^-_2}^2}+&\frac{6}{c} \left(\frac{d S}{d {x^-_2}}\right)^2=\frac{h_{--}(0,x_2^-)}{4G_N}\nonumber\\
&+\frac{1}{16 G_N}\int_{x^-_1}^{x^-_2} dx^- \frac{\left(x^--x^-_1\right)^3 z^2}{\left(x^-_2-x^-\right) \left(x^-_2-x^-_1\right)^3 } \left[\partial_z^2+\frac{3}{z}\partial_z \right]h_{--}(z,x^-)\ .
\end{align}
We can rewrite the above equation in a more transparent way by using Einstein's equation
\begin{align}\label{eqn14}
\frac{d^2 S}{d {x^-_2}^2}+&\frac{6}{c} \left(\frac{d S}{d {x^-_2}}\right)^2=2\pi T_{--}(x_2^-)-\pi\int_{x^-_1}^{x^-_2} dx^- \frac{\left(x^--x^-_1\right)^3 }{\left(x^-_2-x^-\right) \left(x^-_2-x^-_1\right)^3 }T^{bulk}_{--}(z(x^-),x^-)\ ,
\end{align}
where, $T^{bulk}_{--}$ is the bulk stress tensor. We have checked that this equation agrees with direct computations of both sides in a perturbative planer shockwave state.\\

\noindent \textbf{Bulk matter and non-saturation of the QNEC}

In (\ref{eqn14}) it is obvious that the QNEC is saturated when $T^{bulk}_{--}$ vanishes along the geodesic. But in general this is not true. For example, for a shockwave state generated by a null source in the bulk, the QNEC is not saturated when the interval is chosen such that the RT surface intersects with the trajectory of the bulk source. Although the RHS of (\ref{eq:2dQNEC}) does transform like the stress tensor under Virasoro transformations, we have to conclude that in 2d, the holomorphic stress tensor $T$ is not given by the derivatives of the entanglement entropy. 

This feature can be understood from kinematics. Intuitively, $AdS_3/CFT_2$ is different from higher dimensional cases because the RT surface is 1-dimensional, so there is not enough room for the effect of the bulk matter to diffuse on the RT surface. 

More specifically, it is generically possible to change bulk matter configurations, and hence the area of the RT surface, without changing $\langle T_{--}\rangle$ on the boundary. Familiar examples include stars and Schwarzschild blackhole with the same mass, as well as shockwaves generated by a null bulk matter geodesic hovering at different depths. Given that the QNEC is saturated in the vacuum, extra contributions from bulk matter have the potential to make it not saturated.

In higher dimensions, we already know that this does not happen. The reason is easy to understand: only the $\delta$-function term in (\ref{SVar}) is relevant to the higher dimensional QNEC. But in a local gravitational theory, the effect of matter in the bulk will be diffused from the boundary point of view and may not lead to such a local term. So bulk matter is expected to contribute only to $S_{od}$. This is explicitly the case in our results (\ref{S''}) and (\ref{Sod}).

In 2d, bulk matter will still contribute to $S$. But since the entangling cut on the boundary is zero dimensional, there is no longer the difference between $S^{\prime\prime}$ and $S_{od}$. Therefore QNEC is generically not saturated.\\

\noindent \textbf{Bulk energy conditions}

For the QNEC to be satisfied, the bulk source term in (\ref{eqn14}) must have the correct sign. This imposes new energy conditions on admissible bulk matter:
\be\label{anec2}
\int_{x^-_1}^{x^-_2} dx^- \frac{\left(x^--x^-_1\right)^3 }{\left(x^-_2-x^-\right) \left(x^-_2-x^-_1\right)^3 }T^{bulk}_{--}(z(x^-),x^-) \ge 0\ .
\ee
Note that this inequality is not symmetric in $x_1$ and $x_2$ because we applied the QNEC at the point $x_2$. However, there is nothing special about the point $x_2$ and we could apply the QNEC at point $x_1$. That would lead to another inequality exactly like (\ref{anec2}) but with $x_1$ and $x_2$ exchanged. Note that although apparently similar, (\ref{anec2}) is different from the class of energy conditions proved in~\cite{Lashkari:2014kda, Bhattacharya:2014vja} using strong subadditivity. In particular the result of~\cite{Lashkari:2014kda, Bhattacharya:2014vja} is symmetric in exchanging the end points and constrains integrals of the null energy pointing to a spatial direction orthogonal to ours. 

This bulk energy condition (\ref{anec2}) is weaker than the null energy condition, however, stronger than the averaged null energy condition 
\be\label{anec1}
\int_{-\infty}^{\infty} dx^- T^{bulk}_{--}(z,x^-) \ge 0\ .
\ee
In particular, if we send the end points to $\pm\infty$ then we recover the bulk ANEC. To take this limit, we first assume $x_{2}^{-}=-x_{1}^{-}=A>0$ and $x_{2}^{+}=-x_{1}^{+}=\frac{\eta}{A}$. We then multiply both sides of (\ref{anec2}) by A. Finally we take $A\rightarrow\infty$. The corresponding geodesic is a null line on the $-$ direction hovering at a fixed $z$ determined by $\eta$.

\section{Summary and future directions}
In this work we investigated the QNEC with explicit computations in holographic CFTs. We find that in $d\ge3$, nonperturbative smeared shockwave states as well as all states perturbatively close to the vacuum saturate the QNEC when the region is chosen to be a half space. In $d=2$, we find that the QNEC is saturated as long as the associated bulk RT surface does not intersect with $T^{bulk}_{--}$. Otherwise it is in general not saturated. 

In $d\ge3$, we worked out the off-diagonal piece of the second null derivative of the entanglement entropy. We demonstrate that this quantity is directly related to $T^{bulk}_{--}$. Strong subadditivity fixes the sign of this piece and leads to new bulk energy conditions. Our result provides an explicit check for JLMS~\cite{Jafferis:2015del}. We also discussed the relation between bulk matter and boundary relative entropy. In $d=2$, our results together with the QNEC leads to new energy conditions that reduce to the bulk ANEC in a particular limit. In future work, it would be interesting to elucidate the relation between the gap to saturation and the boundary relative entropy and strong subadditivity.

It would also be interesting to ask whether QNEC saturation holds in more general quantum field theories. For theories in $d\ge3$, the techniques developed in~\cite{Balakrishnan:2017bjg} would be useful. In $d=2$, it might be useful to replace the operators in (\ref{eq:HoloS}) by twist operators (and the $\log$ by the standard procedure of taking $n\rightarrow1$ in the replica trick). 

It would also be very interesting to understand the role of bulk matter in AdS/CFT. In particular, we have shown that bulk matter plays an important part in the off-diagonal derivatives of the entanglement entropy in $d\ge3$ and in the relation between energy and entanglement in $d=2$. Relations like (\ref{Sod}) or (\ref{2dQNECNon}) are interesting beyond the QNEC. For example, they may provide non-trivial checks for bulk reconstruction algorithms.

It would be extremely interesting if one can, under some assumptions, invert the corresponding formula (\ref{Sod}) or (\ref{2dQNECNon}) explicitly with, e.g., the inverse Radon transform. Along the lines of~\cite{Lin:2014hva} this would express $T^{bulk}_{--}$ directly as a function of boundary variables such as $S_{EE}$ and the stress tensor. 

Another interesting direction is to use the boundary QNEC to prove (\ref{anec2}), or the bulk ANEC (\ref{anec1}) for more general states or more general holographic theories. 


\section*{Acknowledgments} 
 
We would like to thank Ibou Bah, Netta Engelhardt, Tom Faulkner, Tom Hartman, Jared Kaplan, Liam Fitzpatrick, Silviu Pufu, Edgar Shagoulian, Huajia Wang, and Yifan Wang for discussions. We thank Tom Hartman and Jared Kaplan for comments on the draft. We would like to give a special thanks to Wilke Van Der Schee. The discussion with Wilke about \cite{Ecker:2017jdw} partly motivated this work. ZK acknowledges support from the DARPA, YFA Grant D15AP00108.
DL would like to thank Princeton university and Cornell university for hospitality while some of this work was completed.
DL was supported in part by the Simons Collaboration Grant on the Non-Perturbative Bootstrap.  

\end{spacing}
\bibliographystyle{utphys}
\bibliography{references}

\end{document}